\documentclass[letterpaper]{article}

\PassOptionsToPackage{numbers}{natbib}

\usepackage[final]{nips_2016}
\usepackage[utf8]{inputenc}
\usepackage[T1]{fontenc}
\usepackage{hyperref}
\usepackage{url}
\usepackage{booktabs}
\usepackage{amsfonts}
\usepackage{amsmath}
\usepackage{nicefrac}
\usepackage{microtype}
\usepackage{graphicx}

\graphicspath{{./}}

\DeclareMathOperator*{\argmin}{arg\,min}
\newtheorem{definition}{Definition}
\newtheorem{theorem}{Theorem}

\title{Understanding Convolutional Neural Networks}

\author{
    Jayanth Koushik \\
    Language Technologies Institute \\
    Carnegie Mellon University \\
    Pittsburgh, PA 15213 \\
    \texttt{jkoushik@cs.cmu.edu} \\
}

\begin{document}

\maketitle

\begin{abstract}
    Convoulutional Neural Networks (CNNs) exhibit extraordinary performance
    on a variety of machine learning tasks. However, their mathematical
    properties and behavior are quite poorly understood. There is some
    work, in the form of a framework, for analyzing the operations that they
    perform. The goal of this project is to present key results from this
    theory, and provide intuition for why CNNs work.
\end{abstract}

\section{Introduction}
\subsection{The supervised learning problem}
We begin by formalizing the supervised learning problem which CNNs are
designed to solve. We will consider both regression and classification, but
restrict the label (dependent variable) to be univariate.
Let $X\in\mathcal{X}\subset\mathbb{R}^d$ and $Y\in\mathcal{Y}\subset\mathbb{R}$
be two random variables. We typically have $Y = f(X)$ for some unknown $f$.
Given a sample $\{(x_i,y_i)\}_{i=1,\dots,n}$
drawn from the joint distribution of $X$ and $Y$, the goal of supervised
learning is to learn a mapping $\hat{f}:\mathcal{X}\to\mathcal{Y}$ which
minimizes the expected loss, as defined by a suitable loss function
$L:\mathcal{Y}\times\mathcal{Y}\to\mathbb{R}$. 
However, minimizing over the set of all functions from $\mathcal{X}$ to
$\mathcal{Y}$ is ill-posed, so we restrict the space of hypotheses to some set
$\mathcal{F}$, and define
\begin{align}
	\label{eq:supl}
    \hat{f} = \argmin_{f\in\mathcal{F}}\mathrm{E}[L(Y,f(X))]
\end{align}

\subsection{Linearization}
A common strategy for learning classifiers, and the one employed by kernel
methods, is to linearize the variations in $f$ with a feature representation.
A feature representation is any transformation of the input variable $X$; a
change of variable. Let this transformation be given by $\Phi(X)$. Note that
the transformed variable need not have a lower dimension than $X$. We would
like to construct a feature representation such that $f$ is linearly separable
in the transformed space i.e.
\begin{align}
    f(X) = \langle\Phi(X),w\rangle
\end{align}
for regression, or
\begin{align}
    f(X) = \mathit{sign}(\langle\Phi(X),w\rangle)
\end{align}
for binary classification\footnote{Multi-class classification problems can
be considered as multiple binary classification problems.}. Classification
algorithms like Support Vector Machines (SVM)~\cite{cortes1995support} use a
fixed feature representation that may, for instance, be defined by a kernel.

\subsection{Symmetries}
The transformation induced by kernel methods do not always linearize $f$
especially in the case of natural image classification. To find suitable
feature transformations for natural images, we must consider their invariance
properties. Natural images show a wide range of invariances e.g. to pose,
lighting, scale. To learn good feature representations, we must suppress these
intra-class variations, while at the same time maintaining inter-class
variations. This notion is formalized with the concept of symmetries as
defined next.

\begin{definition}[Global Symmetry]
Let $g$ be an operator from $\mathcal{X}$ to $\mathcal{X}$. $g$ is a
global symmetry of $f$ if $f(g.x) = f(x)\ \forall x \in \mathcal{X}$.
\end{definition}

\begin{definition}[Local Symmetry]
Let $G$ be a group of operators from $\mathcal{X}$ to $\mathcal{X}$ with norm
$|.|$. $G$ is a
group of local symmetries of $f$ if for each $x \in \mathcal{X}$, there exists
some $C_x > 0$ such that $f(g.x) = f(x)$ for all $g \in G$ such that
$|g| < C_x$.
\end{definition}

Global symmetries rarely exist in real images, so we can try to construct
features that linearize $f$ along local symmetries. The symmetries we will
consider are translations and diffeomorphisms, which are discussed next.

\subsection{Translations and Diffeomorphisms}
Given a signal $x$, we can interpolate its dimensions and define $x(u)$ for all
$u \in \mathbb{R}^n$ ($n = 2$ for images). A translation is an operator $g$
given by $g.x(u) = x(u - g)$. A diffeomorphism is a deformation; small
diffeomorphisms can be written as $g.x(u) = x(u - g(u))$.

We seek feature transformations $\Phi$ which linearize the action of local
translations and diffeomorphisms. This can be expressed in terms of a
Lipschitz continuity condition.
\begin{align}
    \label{eq:lips}
    \|\Phi(g.x) - \Phi(x)\| \le C|g|\| x\|
\end{align}

\subsection{Convolutional Neural Networks}
\label{subsec:cnn}
\begin{figure}[!t]
    \includegraphics[width=\textwidth]{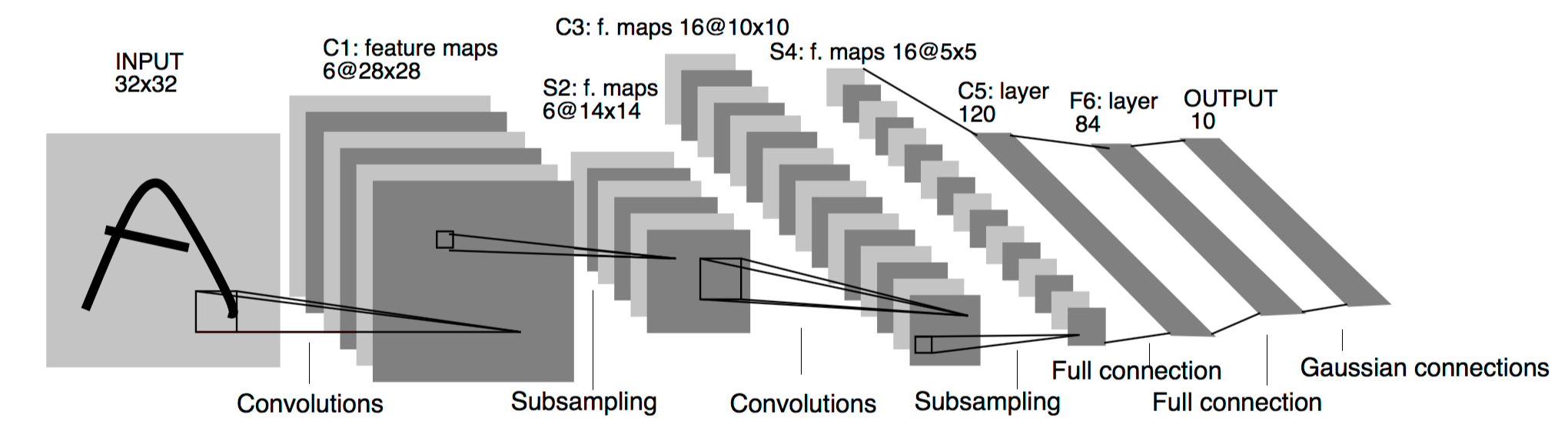}
    \caption{Architecture of a Convolutional Neural Network (from
        \citeauthor{lecun1998gradient}~\cite{lecun1998gradient})}
    \label{figure:cnn}
\end{figure}

Convolutional Neural Networks (CNNs), introduced by~\citet{le1990handwritten}
are a class of biologically inspired neural networks which solve equation
$\eqref{eq:supl}$ by passing $X$ through a series of convolutional
filters and simple non-linearities. They have shown remarkable results in
a wide variety of machine learning problems~\cite{lecun2015deep}. Figure
$\ref{figure:cnn}$ shows a typical CNN architecture.

A convolutional neural network has a hierarchical architecture. Starting
from the input signal $x$, each subsequent layer $x_j$ is computed as
\begin{align}
    x_j = \rho W_j x_{j-1}
\end{align}
Here $W_j$ is a linear operator and $\rho$ is a non-linearity. Typically,
in a CNN, $W_j$ is a convolution, and $\rho$ is a rectifier $\max(x, 0)$ or
sigmoid $\nicefrac{1}{1+\exp(-x)}$. It is easier to think of the operator
$W_j$ as a stack of convolutional filters. So the layers are filter maps and
each layer can be written as a sum of convolutions of the previous layer.

\begin{align}
    x_j(u, k_j) = \rho\Big(\sum_k (x_{j-1}(.,k)\ast W_{j,k_j}(.,k))(u)\Big)
\end{align}

Here $\ast$ is the discrete convolution operator:

\begin{align}
    (f \ast g)(x) = \sum_{u=-\infty}^{\infty} f(u)g(x - u)
\end{align}

The optimization problem defined by a convolutional neural network is
highly non-convex. So typically, the weights $W_j$ are learned by stochastic
gradient descent, using the backpropagation algorithm to compute gradients.

\subsection{A mathematical framework for CNNs}
\citet{mallat2016understanding} introduced a mathematical framework for
analyzing the properties of convolutional networks. The theory is based on
extensive prior work on wavelet scattering (see for example~
\cite{bruna2013invariant,anden2014deep}) and illustrates that
to compute invariants, we must separate variations of $X$ at different
scales with a wavelet transform. The theory is a first step towards
understanding general classes of CNNs, and this paper presents its key
concepts.

\section{The need for wavelets}
Although the framework based on wavelet transforms is quite successful
in analyzing the operations of CNNs, the motivation or need for wavelets
is not immediately obvious. So we will first consider the more general
problem of signal processing, and study the need for wavelet transforms.

In what follows, we will consider a function $f(t)$ where $t\in\mathbb{R}$
can be considered as representing time, which makes $f$ a time varying
function like an audio signal. The concepts, however, extend quite naturally
to images as well, when we change $t$ to a two dimensional vector.
Given such a signal, we are often interested in studying its variations
across time. With the image metaphor, this corresponds to studying
the variations in different parts of the image. We will consider a progression 
of tools for analyzing such variations. Most of the following material is from
the book by \citeauthor{gerald1994friendly}~\cite{gerald1994friendly}.

\subsection{Fourier transform}
The Fourier transform of $f$ is defined as

\begin{align}
	\label{eq:fourier}
	\hat{f}(\omega) \equiv \int_{-\infty}^{\infty}f(t)e^{-2\pi i\omega t}dt
\end{align}

The Fourier transform is a powerful tool which decomposes $f$ into the
frequencies that make it up. However, it should be quite clear from equation
$\eqref{eq:fourier}$ that it is useless for the task we are interested in.
Since the integral is from $-\infty$ to $\infty$, $\hat{f}$ is an average
over all time and does not have any local information.

\subsection{Windowed Fourier transform}
To avoid the loss of information that comes from integrating over all time,
we might use a weight function that localizes $f$ in time. Without going
into specifics, let us consider some function $g$ supported on
$[-T, 0]$ and define the windowed Fourier transform (WFT) as

\begin{align}
	\label{eq:wft}
	\tilde{f}(\omega,t) \equiv \int_{-\infty}^{\infty}f(u)g(u-t)
		e^{-2\pi i\omega u}du
\end{align}

It should be intuitively clear that the WFT can capture local variations
in a time window of width $T$. Further, it can be shown that the WFT also
provides accurate information about $f$ in a frequency band of some width
$\Omega$. So does the WFT solve our problem? Unfortunately not; and this is
a consequence of Theorem $\ref{thm:uncertain}$ which is stated very
informally next.

\begin{theorem}[Uncertainty Principle]
	\label{thm:uncertain}
	\footnote{Contrary to popular belief, the Uncertainty Principle is a
		mathematical, not physical property.}
	Let $f$ be a function which is small outside a time-interval of length $T$,
	and let its Fourier transform be small outside a frequency-band of width
	$\Omega$. There exists a positive constant $c$ such that
		$$ \Omega T \ge c $$
\end{theorem}

Because of the Uncertainty Principle, $T$ and $\Omega$ cannot both be
small. Roughly speaking, this implies that the WFT cannot capture small
variations in a small time window (or in the case of images, a small patch).

\subsection{Continuous wavelet transform}
The WFT fails because it introduces scale (the width of the window) into
the analysis. The continuous wavelet transform involves scale too, but it
considers all possible scalings and avoids the problem faced by the WFT.
Again, we begin with a window function $\psi$ (supported on $[-T,0]$),
this time called a mother wavelet. For some fixed $p\ge 0$, we define
\begin{align}
	\label{eq:wavelet}
	\psi_s(u) \equiv |s|^{-p}\psi\Big(\frac{u}{s}\Big)
\end{align}
The scale $s$ is allowed to be any non-zero real number. With this family
of wavelets, we define the continuous wavelet transform (CWT) as
\begin{align}
	\label{eq:cwt}
    \tilde{f}(s, t) \equiv (f \ast \psi_s)(t)
\end{align}
where $\ast$ is the continuous convolution operator:

\begin{align}
    \label{eq:conv}
    (p \ast q)(x) \equiv \int_{-\infty}^\infty p(u)q(x - u)du
\end{align}

The continuous wavelet transform captures variations in $f$ at a particular
scale. It provides the foundation for the operation of CNNs, as will be
explored next.

\section{Scale separation with wavelets}
Having motivated the need for a wavelet transform, we will now construct
a feature representation using the wavelet transform. Note that convolutional
neural network are covariant to translations because they use convolutions
for linear operators. So we will focus on transformations that linearize
diffeomorphisms.

\begin{theorem}
\label{thm:scale}
Let $\phi_J(u) = 2^{-nJ}\phi(2^{-J}u)$ be an averaging kernel with
$\int\phi(u)du = 1$. Here $n$ is the dimension of the index in $\mathcal{X}$,
for example, $n = 2$ for images. Let $\{\psi_{k}\}_{k=1}^K$ be a set of $K$
wavelets with zero average: $\int\psi_k(u)du = 0$, and from them define
$\psi_{j,k}(u) \equiv 2^{-jn}\psi_k(2^{-j}u)$. Let $\Phi_J$ be a feature
transformation defined as

\begin{align*}
    \Phi_Jx(u,j,k) = |x \ast \psi_{j,k}| \ast \phi_J(u)
\end{align*}

Then $\Phi_J$ is locally invariant to translations at scale $2^J$, and
Lipschitz continuous to the actions of diffemorphisms as defined by equation
$\eqref{eq:lips}$ under the following diffeomorphism norm.

\begin{align}
    \label{eqn:diffnorm}
    |g| = 2^{-J}\sup_{u\in\mathbb{R}^n}|g(u)| +
        \sup_{u\in\mathbb{R}^n}|\nabla g(u)|
\end{align}
\end{theorem}

Theorem $\ref{thm:scale}$ shows that $\Phi_J$ satisfies the regularity
conditions which we seek. However, it leads to a loss of information due
to the averaging with $\phi_J$. The lost information is recovered by
a hierarchy of wavelet decompositions as discussed next.

\section{Scattering Transform}
\begin{figure}[!t]
    \includegraphics[width=\textwidth]{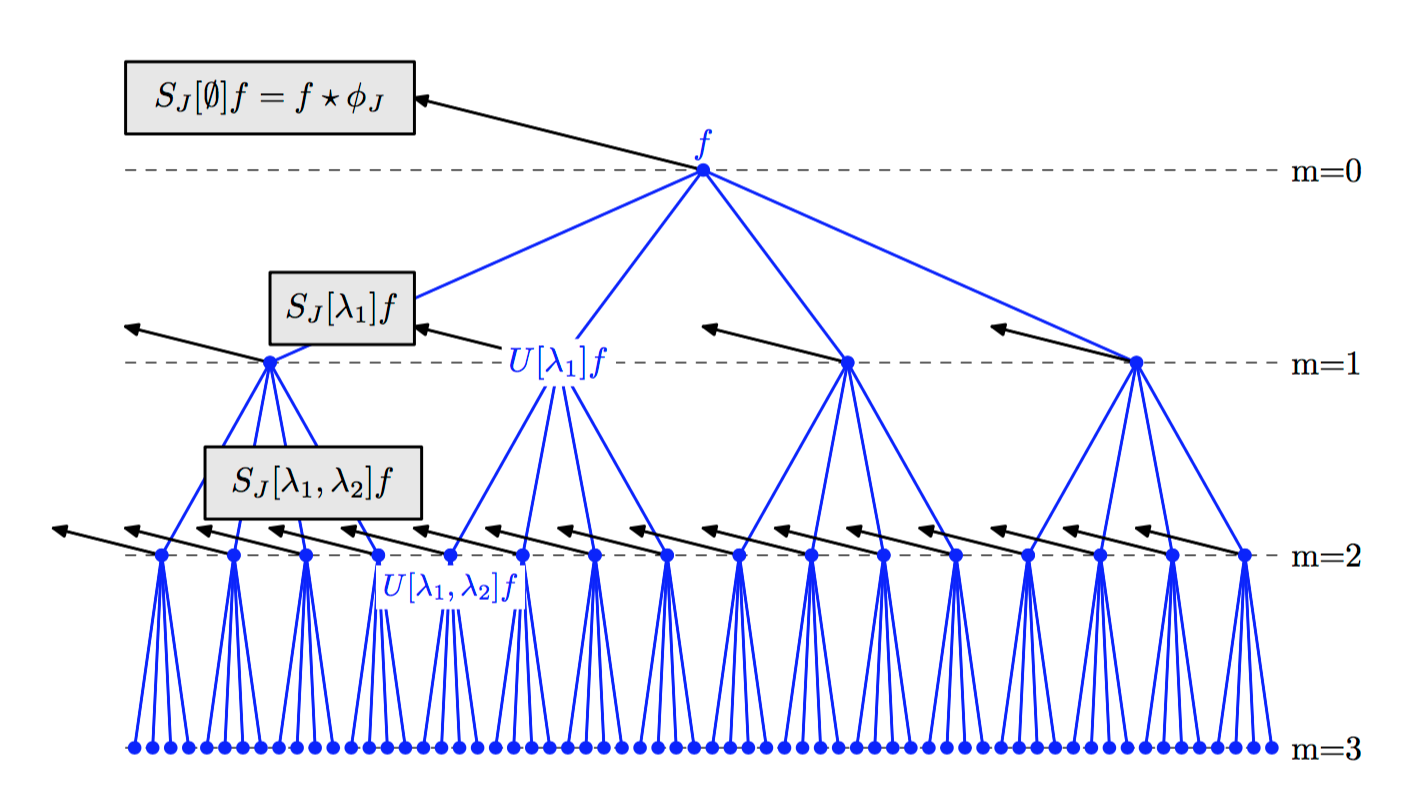}
    \caption{Architecture of the scattering transform (from
        \citeauthor{estrach2012scattering}~\cite{estrach2012scattering})}
    \label{figure:scatter}
\end{figure}

Convolutional Neural Networks transform their input with a series of linear
operators and point-wise non-linearities. To study their properties, we first
consider a simpler feature transformation, the scattering transform introduced
by \citeauthor{mallat2012group}~\cite{mallat2012group}. As was discussed in
section $\ref{subsec:cnn}$, CNNs compute multiple convolutions across channels
in each layer; So as a simplification, we consider the transformation obtained
by convolving a single channel:

\begin{align}
    x_j(u, k_j) = \rho\Big((x_{j-1}(., k_{j-1}) \ast W_{j, h})(u)\Big)
\end{align}

Here $k_j = (k_{j-1}, h)$ and $h$ controls the hierarchical structure of the
transformation. Specifically, we can recursively expand the above equation
to write

\begin{align}
    x_J(u, k_J) = \rho(\rho(\dots\rho(x\ast W_{1,h_1})\ast\dots)\ast W_{J,h_J})
\end{align}

This produces a hierarchical transformation with a tree structure rather than
a full network. It is possible to show that the above transformation has
an equivalent representation through wavelet filters i.e. there exists a
sequence $p \equiv (\lambda_{1},\dots,\lambda_{m})$ such that

\begin{align}
    \label{eqn:scatter}
    x_J(u, k_J) = S_J[p]x(u) \equiv (U[p]x \ast \phi_J)(u) \equiv
        (\rho(\rho(\dots\rho(x\ast \psi_{\lambda_1})\ast\dots)\ast
        \psi_{\lambda_m})\ast\phi_J)(u)
\end{align}

where the $\psi_{\lambda_i}$s are suitably choses wavelet filters and $\phi_J$
is the averaging filter defined in Theorem $\ref{thm:scale}$. This is the
wavelet scattering transfom; its structure is similar to that of a
convolutional neural network as shown in figure $\ref{figure:scatter}$, but its
filters are defined by fixed wavelet functions instead of being learned from
the data. Further, we have the following theorem about the scattering
transform.

\begin{theorem}
    \label{thm:scatterdiff}
    Let $S_J[p]$ be the scattering transform as defined by equation
    $\eqref{eqn:scatter}$. Then there exists $C > 0$ such that for all
    diffeomorphisms $g$, and all $L^2(\mathbb{R}^n)$ signals $x$,
    \begin{align}
        \|S_J[p]g.x - S_J[p]x\| \le Cm|g|\|x\|
    \end{align}

    with the diffeomorphism norm $|g|$ given by equation
    $\eqref{eqn:diffnorm}$.
\end{theorem}

Theorem $\ref{thm:scatterdiff}$ shows that the scattering transform is
Lipschitz continuous to the action of diffemorphisms. So the action of small
deformations is linearized over scattering coefficients. Further, because of
its structure, it is naturally locally invariant to translations. It has
several other desirable properties~\cite{estrach2012scattering}, and can be
used to achieve state of the art classification errors on the MNIST digits
dataset~\cite{bruna2013invariant}.

\section{General Convolutional Neural Network Architectures}
The scattering transform described in the previous section provides a simple
view of a general convolutional neural netowrk. While it provides
intuition behind the working of CNNs, the transformation suffers from high
variance and loss of information because we only consider single channel
convolutions. To analyze the properties of general CNN architectures, we
must allow for channel combinations. \citeauthor{mallat2016understanding}
\cite{mallat2016understanding} extends previously introduced tools to develop
a mathematical framework for this analysis. The theory is, however, out of the
scope of this paper. At a high level, the extension is achieved by replacing
the requirement of contractions and invariants to translations by contractions
along \emph{adaptive} groups of local symmetries. Further, the wavelets are
replaced by adapted filter weights similar to deep learning models.

\section{Conclusion}
In this paper, we tried to analyze the properties of convolutional neural
networks. A simplified model, the scattering transform was introduced as
a first step towards understanding CNN operations. We saw that
the feature transformation is built on top of wavelet transforms which
separate variations at different scales using a wavelet transform. The
analysis of general CNN architectures was not considered in this paper, but
even this analysis is only a first step towards a full mathematical
understanding of convolutional neural networks.

\bibliographystyle{plainnat}
\bibliography{references}

\end{document}